\documentclass[twocolumn,prd,showpacs,10pt]{revtex4}
\usepackage{graphicx}
\usepackage{dcolumn}
\usepackage{bm}

\newcommand{\square}{\kern1pt\vbox{\hrule height 1.2pt\hbox{\vrule
width 1.2pt\hskip 3pt
\vbox{\vskip 6pt}\hskip 3pt\vrule width 0.6pt}\hrule
height 0.6pt}\kern1pt}
\newcommand{\beq}{\begin{equation}}
\newcommand{\beqn}{\begin{eqnarray}}
\newcommand{\eeq}{\end{equation}}
\newcommand{\eeqn}{\end{eqnarray}}

\begin{document}

\title{An Observational Test of Holographic Inflation}

\author{James E. Lidsey and David Seery}
\affiliation{Astronomy Unit, School of Mathematical Sciences, 
Queen Mary, University of London, Mile End Road,
London, E1 4NS, UK}

\begin{abstract}
Observational consequences of inflationary cosmology 
in the holographic dual of the Randall-Sundrum type II 
braneworld scenario, as motivated by the AdS/CFT correspondence,
are investigated. High energy corrections to the standard 
four-dimensional Friedmann equation induce 
a corresponding modification to the form of 
the single-field inflationary consistency 
equation based on Einstein gravity. The degree of departure from the 
standard expression is determined by the ratio, $r$, of the primordial 
tensor and scalar perturbation amplitudes and the coefficient, $c$, 
of the conformal anomaly in the dual gauge theory.  
It is found that a necessary condition for detecting 
such a correction with the next generation of cosmic microwave
background (CMB) polarization experiments is that $r \ge 0.06$. The 
bound is tightened to $r>0.3$ for values
of the central charge that are compatible with known compactifications 
of type IIB string theory as parametrized in terms of F-theory
compactification on Calabi-Yau four-folds.
This is close to the present upper bounds inferred 
from combined observations of the CMB anisotropy power spectrum 
and high redshift surveys. We conclude that if 
such modifications to the inflationary consistency equation are to be 
observable, the gravitational wave background should be 
detected in the near future. A further consequence 
of the non-standard dynamics at high energies is that the initial 
state of the universe is a quiescent singularity with a finite 
density and pressure.  
\end{abstract}

\vskip 1pc 
\maketitle

\section{Introduction}

The AdS/CFT correspondence represents one of the most striking 
insights to have emerged from string/M- theory in the past decade 
\cite{adscft}. 
(For a review, see, \cite{adsreview}). This correspondence implies 
that a gravitational theory on $(d+1)$--dimensional 
Anti-de Sitter $({\rm AdS}_{d+1})$ space admits a dual description 
in terms of a conformal field theory (CFT) propagating on the $d$--dimensional 
boundary. 

On the other hand, 
recent years have witnessed tremendous advances in observational cosmology 
with the high--precision data announced by the 
Wilkinson Microwave Anisotropy Probe (WMAP) and 
other Cosmic Microwave Background (CMB) surveys 
strongly supporting the core predictions of inflationary 
cosmology, namely a spatially--flat universe 
with a primordial spectrum of adiabatic, gaussian and nearly scale--invariant
density perturbations \cite{WMAP,seljak}. 

In view of these developments, there is a pressing need to 
understand how inflation may be incorporated within string/M-theory
and to confront the predictions of the models with present-day and forthcoming 
cosmological observations. Motivated by developments in string/M-theory, the 
braneworld paradigm -- where our observable universe is viewed as 
a co-dimension one brane propagating in a higher-dimensional `bulk' space -- 
has received considerable attention. (See \cite{branereview} for reviews). 
Of particular interest is the 
Randall-Sundrum type II (RSII) scenario, where the brane is embedded in 
an Einstein space sourced by a negative 
cosmological constant \cite{RSbrane}. The effective localization of 
the graviton zero-mode on the brane can be understood within the 
AdS/CFT context by interpreting the model as a cut--off, strongly coupled 
conformal gauge theory coupled to four--dimensional 
Einstein gravity \cite{brave,gubser}. This holographic dual 
formulation of the RSII scenario 
can be viewed as a four-dimensional theory in 
its own right with an effective action comprised of 
a sum of contributions, $S_{\rm dual} = S_{\rm loc} + \Gamma + S_{\rm mat}$, 
where $S_{\rm loc}$ represents the Einstein-Hilbert action 
(including a possible cosmological constant), $S_{\rm mat}$ represents 
standard matter fields and $\Gamma$ is the action for the 
conformal anomaly of the gauge theory 
\cite{brave,gubser,noz,kiritsis,difficultcalc,boer}. In 
the spatially isotropic Friedmann-Robertson-Walker (FRW) cosmology, 
such a dual interpretation modifies the form of the Friedmann equation
away from the standard $H^2 \propto \rho$ dependence on the energy 
density.  

Here we investigate whether such modifications to the cosmic 
dynamics may leave an observable signature in the CMB. As well as 
providing a causal mechanism for generating the primordial 
spectrum of density perturbations, inflation also produces 
a spectrum of tensor (gravitational wave) fluctuations 
\cite{gwstar}. A direct detection 
of the amplitude and tilt of such a spectrum would 
yield crucial information on the energy scale of inflation. 
Moreover, in conventional single-field slow-roll inflation, the amplitudes 
of the perturbation spectra, $A^2_{S,T}$, are related to the tensorial 
tilt, $n_T$, through a `consistency' 
equation that is independent of the form of the inflaton 
potential \footnote{We adopt the normalization   
conventions of Ref. \cite{lidseyreview}.
We also define the observable 
tensor-to-scalar ratio in terms of the CMB quadrupole moments 
modified to account for a dark energy component \cite{turnerwhite}. This 
is related to the presently favoured definition used by the WMAP 
team such that $r_{\rm WMAP} = (16/9.6) r=  1.67r$.}: 
\begin{equation}
\label{standconsist}
\frac{A_T^2}{A_S^2} = -\frac{1}{2} n_T  .
\end{equation}
Standard slow-roll inflation driven by a single inflaton
field could therefore be verified or ruled out by confronting 
the relation (\ref{standconsist}) with observations. 
Eq. (\ref{standconsist}) 
also arises as a consistency equation in a number of braneworld inflationary
scenarios \cite{persistency}. 

In general, modifications to the Friedmann 
equation such as those arising in the holographic dual of the RSII 
scenario will induce departures from the standard  
consistency equation (\ref{standconsist}). We determine the prospects 
for detecting such a departure in future CMB polarization experiments. 
Polarization of the CMB represents the most 
promising route toward a direct detection of the 
gravitational wave background, since the divergence-free $B$-mode 
of the polarization anisotropy can only be produced (at leading-order)
by tensor perturbations \cite{Bmode}. Consequently, considerable attention has 
focused recently on understanding the experimental and theoretical 
issues involved in detecting a primordial $B$-mode polarization 
and a number of experiments are planned
for the near future. (See \cite{polarreview} for a review). 
For example, the Planck surveyor \cite{planck} will  
detect a scalar-to-tensor ratio above $r \approx 0.02$, 
while Clover \cite{clover}
will be sensitive to values as small as $r \approx 0.06$.
We find that a departure from the 
standard inflationary consistency equation could in principle 
be detectable if the amplitude of the gravitational wave spectrum 
exceeds $r \approx 0.06$.    

\section{Holographic Inflation}

We begin by summarizing the derivation of the modified form 
of the Friedmann equation in the holographic dual of the RSII scenario
\footnote{The reader is referred to \cite{kiritsis,boer} for 
more detailed discussions.}. 
The dual four-dimensional action can be calculated 
from the holographic renormalization group approach 
developed by de Boer, Verlinde and Verlinde \cite{boer}. This is 
based on the Hamilton-Jacobi (HJ) equation of General Relativity
\cite{salopek}. In this prescription, the fourth-order 
contribution to the generating functional of the HJ equation 
is identified as the effective action, $\Gamma$, for the gauge theory.  
The dual action then takes the form  $S_{\rm dual} = S_{\rm loc} + \Gamma + 
S_{\rm mat}$, where $S_{\rm loc}$ contains terms that 
are no higher than second-order in derivatives and includes the Einstein
action that renders the bulk AdS action finite. (We will 
assume implicitly that the tension of the brane is tuned to cancel 
the local (divergent) contribution arising from an effective 
cosmological constant). The matter fields on the brane are 
described by the action $S_{\rm mat}$. 
Extremizing the variation of 
the dual action with respect to the metric tensor yields 
the gravitational field equations, 
$\hat{m}_P^2 G_{\mu\nu} = T_{\mu\nu} +V_{\mu\nu}$
where $G_{\mu\nu}$ is the Einstein tensor, $T_{\mu\nu} = 2
(-g)^{-1/2} \delta S_{\rm mat}/ \delta g^{\mu\nu}$ is the 
energy-momentum of the ordinary matter sector,  
$V_{\mu\nu} \equiv 2 (-g)^{-1/2} \delta \Gamma /\delta g^{\mu\nu}$ 
is the effective stress tensor of the CFT and 
$\hat{m}_P = m_P/\sqrt{8\pi}$ is the reduced Planck mass. 

The specific form of the 
trace of the CFT energy-momentum tensor 
follows directly from the fourth-order HJ equation 
and is given by \cite{boer,salopek,martelli}
\begin{equation}
\label{traceanomaly}
{V^{\mu}}_{\mu} 
= \frac{\ell^3 M^3}{8} \left( R_{\mu\nu}R^{\mu\nu} - \frac{1}{3}R^2 \right)  ,
\end{equation}
where $\ell$ is the AdS radius of curvature and 
is determined in terms of the five-dimensional bulk 
cosmological constant, $\Lambda$, such that $\Lambda = -12M^3/\ell^2$, 
where $M$ is the five-dimensional Planck scale. 

On the other hand, it is also known from direct calculation 
in four dimensions \cite{difficultcalc} that the trace anomaly has the form 
${V^{\mu}}_{\mu} = - \tilde{a} C_{\mu\nu\lambda\rho} 
C^{\mu\nu\lambda\rho} -c G$, where $C_{\mu\nu\lambda\rho}$ is the 
Weyl tensor, $G$ is the Gauss-Bonnet combination 
of curvature invariants and the constants $\{ \tilde{a} ,c \}$ are determined 
by the field content of the CFT such that  
\begin{eqnarray}
\label{adef}
\tilde{a} = - \frac{1}{120(4\pi )^2} \left( N_S +6N_F +12N_V \right) 
\nonumber \\
c = \frac{1}{360 (4 \pi )^2} \left( N_S +11N_F +62 N_V \right)  ,
\end{eqnarray}
where $N_I$ denotes the number of scalar, fermionic and vector
degrees of freedom, 
respectively, and the fermions are Dirac fermions. The field content 
of Yang-Mills theory yields $N_S=6N^2$, $N_F=2N^2$ and $N_V=N^2$ 
for $N\gg 1$, where $N$ denotes the number of colours in the gauge 
group. This implies that $c= (N/8\pi)^2$ and $\tilde{a} =-c$. 
Hence, equating the above expression 
for the trace with Eq. (\ref{traceanomaly}) 
relates the conformal anomaly coefficient directly to the 
AdS radius, $c=(M \ell /2)^3$. 

For a spatially flat FRW line element with scale factor $a$, 
we may define $V_{00} \equiv \sigma$ and $V_{ij}\equiv \sigma_P 
a^2 \delta_{ij}$. The Bianchi identity, together with 
conservation of the energy-momentum, ${T^{\mu\nu}}_{; \mu} =0$, 
then implies that ${V^{\mu\nu}}_{; \mu} =0$, i.e., that 
\begin{equation}
\label{anomalyconserve}
\dot{\sigma} = - 3H (\sigma +\sigma_P)  ,
\end{equation}
where $H\equiv \dot{a}/{a}$ 
is the Hubble parameter. On the other hand, 
the trace anomaly (\ref{traceanomaly}) simplifies to 
\begin{equation}
\label{simpletrace}
\sigma -3\sigma_P = 24c \frac{\ddot{a}}{a} \frac{\dot{a}^2}{a^2} 
\end{equation}
and substituting Eq. (\ref{simpletrace}) into 
Eq. (\ref{anomalyconserve}) yields, up on integration, the 
solution $\sigma = \chi +6cH^4$, where $\chi \propto 1/a^4$ is an 
effective radiation term. During inflation, this term is rapidly redshifted 
away and we therefore neglect its contribution. The $(00)$-component 
of the field equations then yields the effective Friedmann equation 
\begin{equation}
\label{friedmann}
H^2 = \frac{\hat{m}_P^2}{4c} \left[ 1+ \epsilon \sqrt{1-
\frac{8c}{3\hat{m}_P^4} \rho } \right] ,
\end{equation}
where $\epsilon = \pm 1$. 
Note that there exists an {\em upper bound} to the energy density, 
$\rho < \rho_{\rm max} = 3\hat{m}_P^4/(8c)$.   
The $\epsilon =-1$ branch reduces to the 
correct form of the Friedmann equation in the low-energy limit, 
$H^2 \propto \rho$, and we therefore focus on this case in the following. 

We will assume the dynamics of the universe is driven by a single
scalar field with self-interaction potential $V(\phi)$. 
Covariant conservation of its energy-momentum tensor therefore implies that
\begin{equation}
\label{scalareom}
\ddot{\phi} + 3H\dot{\phi} + \frac{dV}{d\phi} =0  .
\end{equation}
Eq. (\ref{scalareom}) may also be expressed in the 
form $\dot{\rho}=-3H\dot{\phi}^2$, where $\rho = \dot{\phi}^2 /2 +V$ 
represents the energy density of the field.  

It proves convenient to define a new variable, $\theta$: 
\begin{equation}
\rho \equiv \frac{3\hat{m}_P^4}{8c} \cos^2 \theta  ,
\end{equation}
where $\theta \in [0,\pi/2 ]$, and an equation of state 
parameter, $\gamma \equiv \dot{\phi}^2/\rho$. 
Differentiating Eq. (\ref{friedmann}) then implies that 
\begin{equation}
\label{Hdot}
\dot{H}= - \frac{1}{2\hat{m}_P^2} \frac{\gamma \rho}{\sin \theta}  .
\end{equation}
The necessary and sufficient condition for inflation, 
$\dot{H} +H^2 >0$, may then be expressed as a lower 
limit on the value of $\theta$ for a given equation of state parameter: 
\begin{equation}
\label{conforinf}
\sin \theta > \frac{3\gamma}{4-3\gamma} .
\end{equation}

For a semi-positive-definite self-interaction potential, 
$V \ge 0$, the equation of state is bounded, $0\le \gamma \le 2$. 
As the energy density approaches the upper bound 
$\rho \rightarrow \rho_{\rm max}$ $(\theta \rightarrow 0$), it follows that
the pressure, Hubble parameter and scale 
factor of the universe also remain finite. 
However, Eq. (\ref{Hdot}) implies that 
the universe has infinite deceleration at this point,  
$\ddot{a} \rightarrow -\infty$. The model is therefore geodesically 
incomplete, since the Ricci scalar diverges. On the other hand, 
this does not represent a conventional big bang singularity 
since the energy density remains finite. Such a past `quiescent'
curvature singularity was also recently uncovered in a 
particular class of five-dimensional braneworld models \cite{brown}. 
It is similar to the sudden future singularities 
discussed recently by Barrow \cite{barrow}, although the pressure of the fluid 
is also divergent in Barrow's cases. In effect, the existence of this 
quiescent singularity delays the onset of inflationary expansion. 
Indeed, a necessary (but not sufficient) condition for inflation is 
that $\gamma <2/3$, whereas this condition is necessary {\em and} 
sufficient in the standard scenario. 

We may view Eqs. (\ref{friedmann})--(\ref{scalareom}) as an effective 
four-dimensional cosmology. 
Conservation of energy-momentum, Eq. (\ref{scalareom}), then 
implies that the adiabatic curvature perturbation  
on a uniform density hypersurface, $\zeta = H\delta \phi 
/\dot{\phi}$, is conserved on large scales \cite{wands}.
(We assume implicitly throughout that scales that are observable today
first crossed the Hubble radius during a phase of slow-roll inflation, 
where $|\dot{H}|/H^2 \ll 1$ and $|\ddot{\phi}| \ll H|\dot{\phi}|$). 
As a result, the amplitude of a perturbation mode when it re-enters the 
Hubble radius after inflation is given by $A_S^2 = 4 \langle
\zeta^2 \rangle /25$, where the right-hand side is evaluated when the mode 
first crosses the Hubble radius during inflation, i.e., when $k=aH$, 
where $k$ is the comoving wavenumber. The scalar field fluctuation 
at this epoch is determined by the Gibbons-Hawking temperature 
of de Sitter space, $\langle \delta \phi \rangle = H^2/(4\pi^2)$, and 
the scalar perturbation amplitude is therefore given by  
\begin{equation}
\label{scalarperturbs}
A_S^2 = \frac{1}{25\pi^2}\frac{H^4}{\dot{\phi}^2}  .
\end{equation}

The tensor and matter perturbations decouple from one another to first-order. 
The large-scale amplitude of each tensor mode when 
crossing the Hubble radius during inflation is therefore 
determined solely by the expansion rate of the universe,  
as in conventional slow-roll inflation \cite{gwstar}. 
This implies that 
\begin{equation}
\label{tensorperturbs}
A_T^2= \frac{4}{25\pi} \frac{H^2}{m_P^2} = \frac{1}{200 \pi^2 c}
( 1- \sin \theta )  .
\end{equation}
The tilt of the tensor perturbation spectrum, 
$n_T \equiv d \ln A_T^2/d \ln k$, is then determined by differentiating 
Eq. (\ref{tensorperturbs}) with respect to comoving wavenumber and 
substituting in Eqs. (\ref{scalarperturbs}) and (\ref{tensorperturbs}). 
We find that 
\begin{equation}
\label{tensortilt}
n_T= -2 \frac{A_T^2}{A_S^2}
\frac{1}{\sin \theta}  .
\end{equation}

Comparison between Eqs. (\ref{standconsist}) and (\ref{tensortilt}) 
implies that the standard inflationary consistency equation is 
modified due to the corrections arising in the 
Friedmann equation (\ref{friedmann}). It is still the case that 
$n_T \le 0$, implying immediately that this holographic scenario could be 
ruled out by a detection of a positive spectral index. 
More specifically, however, we deduce that for a given 
scalar-to-tensor ratio, the magnitude of the spectral 
index is {\em enhanced} by a factor of $\sin \theta$ relative to the standard 
scenario.

\section{Holographic Consistency Relation}

Eq. (\ref{tensortilt}) can be rearranged into 
a more convenient form by substituting in Eq. (\ref{tensorperturbs}): 
\begin{equation}
\label{consistency}
n_T +2\frac{A_T^2}{A_S^2} = - \frac{400\pi^2 c}{1-
200\pi^2 c A_T^2} \frac{A^4_T}{A_S^2}
\end{equation}
and Eq. (\ref{consistency}) can be rewritten in terms of observable 
parameters by relating the perturbation amplitudes $A^2_{S,T}$ directly 
to the corresponding quadrupole variances in the power 
spectrum of the CMB, $C_2^{S,T}$. We consider a (spatially flat) concordance 
cosmology with a dark energy density parameter 
$\Omega_{\Lambda}=0.66$ and reduced Hubble parameter $h=0.66$. The observable
tensor-to-scalar ratio is then defined by 
$r \equiv T/S$, where $T \equiv 5C_2^T/4\pi = 1.4 f_T
A_T^2$ and $S \equiv 5C_2^S/4\pi = 0.1f_S A_S^2$, and the `transfer
functions' are defined by $f_S= 1.04 -0.82\Omega_{\Lambda}+
2\Omega^2_{\Lambda} \approx 1.37$ and $f_T= 1.0 -0.03 \Omega_{\Lambda}
-0.1 \Omega^2_{\Lambda} \approx 0.94$, respectively \cite{turnerwhite}. 
These definitions take into account the non--negligible contribution to the 
quadrupole from the late-time integrated Sachs-Wolfe 
effect in a dark energy dominated 
universe. The COBE normalization of the temperature anisotropy 
power spectrum is $S=5.5 \times 10^{-11}$ \cite{COBEnorm}
and it follows that $r=T/S = 9.6A_T^2/A_S^2=1.8 \times 10^{10} T$. 
Hence, the consistency equation (\ref{consistency}) 
may be written in the form 
\begin{equation}
\label{equivconsist}
n_T+ \frac{r}{4.8} = - \frac{1}{4.8} r r_l  \, 
\end{equation}
where 
\begin{equation}
\label{defrl}
r_l \equiv \frac{\tilde{c}r}{1-\tilde{c}r} , \qquad 
\tilde{c} \equiv  8.25 \times 10^{-8} c  .
\end{equation}

Since Eq. (\ref{equivconsist}) is independent 
of the specific functional form of the inflaton potential, 
it may be interpreted as the 
consistency equation for single field inflation in the holographic 
dual of the RSII braneworld cosmology. For a given set of observations, the 
only free parameter in this relation 
is the coefficient of the conformal anomaly, $c$, or equivalently, 
the number of colours in the dual gauge theory. 
Since a departure from the standard consistency 
equation is parametrized by a non--zero right-hand side in 
Eq. (\ref{equivconsist}), we will refer to $r_l$ as the 
`departure parameter'.  

The question that now arises, therefore, is whether 
the holographic consistency equation can be observed. 
Song and Knox (SK) \cite{songknox}
have discussed the prospects of detecting a departure 
from the standard consistency equation that has precisely the form 
(\ref{equivconsist}). In SK, however, $r_l$ is viewed as a free parameter 
that arises from the inclusion of quantum loop 
corrections to the two-point correlation function of the inflaton
\footnote{We neglect the possible effects of such field theoretic 
corrections in the present work.}. 
Although the physical interpretation of the departure parameter 
is different in the present context, we may employ the  
results of SK to investigate 
the detectability prospects of the holographic consistency equation.  
SK consider a future CMB polarization experiment with full sky coverage, 
an angular resolution in the range $1.0' \le \varphi \le 30.0'$, 
and a noise level in the range $1\mu {\rm K} \cdot {\rm arcmin} 
< \Delta_T < 15 \mu{\rm K} \cdot {\rm arcmin}$, 
where $\Delta_T= 1/\sqrt{2\omega}$ and $\omega$ 
is the weight per solid angle. They further assume an optical depth 
parameter $\tau = 0.17$ and proceed to calculate the anticipated error on 
$n_T+r/4.8$ as a function of the tensor-to-scalar ratio 
$r$. The results are summarized in Table I
for a $B$-mode that has been `cleaned' by estimating 
the projected lensing potential from the four-point function 
of the temperature and polarization fields \cite{ohu}. 

\begin{figure}[!t]

\includegraphics[width=7.2cm]{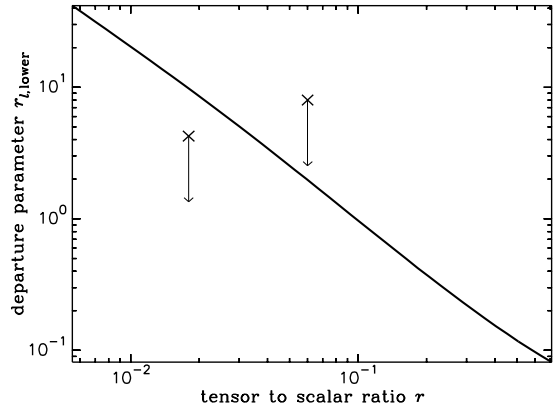}

\caption[] {\label{fig1holography}
Illustrating the minimal value of the departure parameter $r_l$ 
that will lead to a detection of the holographic consistency equation 
by the future full-sky CMB polarization experiment considered in SK 
\cite{songknox}. 
In principle, for a given fiducial value of the tensor-to-scalar ratio, $r$, 
a detection is possible in the region of parameter 
space above the solid line. The crosses denote points 
corresponding to the future full-sky polarization experiment 
considered by VPJ \cite{vpj}, where
the errors on $r$ saturate the upper limit (\ref{upperlimit}). 
For these values of $r$, departures from the standard 
inflationary consistency equation will only be detectable for 
values of $r_l$ smaller than those indicated by the crosses. 
Hence, compatibility between the upper and lower bounds requires $r \ge 0.06$.
}
\end{figure}

In practice, this implies that for a given fiducial  
value of $r$, one may determine how large the departure parameter $r_l$ 
must be for the modification in the holographic consistency equation
to be observable. A necessary condition is that the 
right-hand side of Eq. (\ref{equivconsist}), $Q \equiv rr_l/4.8$, 
should exceed $\sigma_T$, the anticipated error in $n_T+r/4.8$. 
The results are shown in Fig. \ref{fig1holography} and the third 
column of Table I. The lower limit on 
$r_l$ increases monotonically for decreasing $r$.  

\begin{table}
\begin{center}
\begin{tabular}{|c|c|c|c|c|}
\hline 
$r$        & $\sigma_T$ & $r_{l,\rm lower}$   & $c_{\rm min}$         & $N_{\rm min}$       \\ 
\hline \hline
$0.71$      & $0.012$    & $0.08$              & $1.3 \times 10^6$     & $2.8 \times 10^4$   \\
$0.63$      & $0.012$    & $0.09$              & $1.6 \times 10^6$     & $3.2 \times 10^4$    \\
$0.51$      & $0.012$    & $0.12$              & $2.5 \times 10^6$     & $4.0 \times 10^4$    \\
$0.40$      & $0.013$    & $0.15$              & $4.0 \times 10^6$     & $5.0 \times 10^4$    \\
$0.32$      & $0.014$    & $0.20$              & $6.3 \times 10^6$     & $6.3 \times 10^4$    \\
$0.21$      & $0.016$    & $0.36$              & $1.6 \times 10^7$     & $1.0 \times 10^5$    \\
$0.10$      & $0.020$    & $0.92$              & $5.6 \times 10^7$     & $1.9 \times 10^5$    \\
$0.07$     & $0.023$    & $1.47$               & $9.7 \times 10^7$     & $2.5 \times 10^5$    \\
$0.06$     & $0.025$    & $2.00$               & $1.4 \times 10^8$     & $2.9 \times 10^5$    \\  
\hline
$0.05$     & $0.027$    & $2.73$               & $1.9 \times 10^8$     & $3.4 \times 10^5$    \\
$0.02$     & $0.036$    & $9.01$               & $5.7 \times 10^8$     & $6.0 \times 10^5$    \\
\hline 
\end{tabular}
\caption{
Summarizing the prospects for detecting the holographic 
consistency equation (\ref{equivconsist}). The second 
column represents the expected error on $n_T+r/4.8$, denoted by $\sigma_T$, 
for fiducial values of the tensor-to-scalar ratio, $r$,
in the future CMB polarization experiment of SK \cite{songknox}. 
Angular resolution is specified to be $\varphi =1'$ with a noise level 
$\Delta_T = 3 \mu {\rm K} \cdot {\rm arcmin}$. The third column 
$r_{l,\rm lower}$ denotes the minimum value of the departure parameter 
that will lead to an observable departure from the standard consistency 
equation. The fourth and fifth columns represent the corresponding limits 
on the conformal anomaly coefficient, $c$, and the number of 
colours in the gauge theory, $N$, respectively. The horizontal line 
at $r \approx 0.06$ indicates the critical value of $r$ where 
the projected observable errors on $Q \equiv rr_l/4.8$ 
are comparable to the magnitude of $Q$ itself given that 
$r_l> r_{l,\rm lower}$ (see the text for details).  
Detection of the holographic consistency equation is only possible 
for $r \ge 0.06$. 
}
\end{center}
\label{table1}
\end{table}

SK regard $r_l$ as a free parameter
of the theory rather than an observable. However, in the 
holographic consistency relation, the departure parameter depends  
on the tensor-to-scalar ratio and will therefore 
be subject to experimental errors. Although it is to be expected that 
the most significant errors will arise from measurements 
of the tensorial tilt, the error in the departure parameter will
limit how low the observed value of $r$ can be if  
a deviation is to be observable. 
The error in $Q \equiv rr_l/4.8$ (the right-hand side of the 
holographic consistency equation (\ref{equivconsist})) 
is estimated to be 
$|\delta Q| = r_l (2+r_l) |\delta r|/4.8$ and, if a non-zero value of $Q$ 
is to be detectable, a necessary condition is that $|\delta Q| <Q$. 
Hence, given a fiducial value of the tensor-to-scalar ratio with an 
associated error, this constraint may be interpreted as an 
{\em upper limit} on the value of the departure parameter that will 
lead to a detection, i.e, we require that 
\begin{equation}
\label{upperlimit}
r_l < \frac{r}{|\delta r|} -2   .
\end{equation}
Consequently, by 
combining this constraint with the lower limit $r_{l,\rm lower}$ 
shown in Fig. \ref{fig1holography}, 
we may estimate the smallest allowed value of 
$r$ that will lead (in principle) to a detection of the holographic
consistency equation. 

Recently, Verde, Peiris and 
Jimenez (VPJ) have quantified the level of accuracy at which a  
primordial gravitational wave background could be detected 
using a variety of space- and ground-based polarization 
experiments \cite{vpj}. (See also \cite{othergravy}). Specifically, they have 
determined the anticipated errors in the tensor-to-scalar ratio for   
fiducial values of $r$ {\em without} 
imposing the standard consistency equation.  
(In general, fixing $n_T$ through Eq. (\ref{standconsist}) 
greatly reduces the error in $r$).  
For the case of a realistic future all-sky satellite experiment 
with no foreground subtraction or delensing, VPJ conclude that 
an accuracy of $|\delta r |= 0.003$ when $r=0.02$ is possible, 
while a ratio of $r =0.06$ could be detected 
with an accuracy of $|\delta r | = 0.006$. They also 
find that the constraints on $|\delta r|$ do not alter significantly 
even when foreground contamination is reduced to 1$\%$ 
of its original level or when delensing techniques are applied
\cite{vpj}. This level of accuracy at $r=0.02$ 
implies that $r_l< 5$ is required 
for a detection of Eq. (\ref{equivconsist}) and this is inconsistent 
with the results of Fig. \ref{fig1holography}, as illustrated by 
the cross in the diagram. On the other hand, if $r=0.06$, 
we would require $r_l< 8$ and 
this is compatible with Fig. \ref{fig1holography}. 
We may conclude, therefore, that 
a future detection of the holographic consistency equation will require 
the tensor-to-scalar ratio to satisfy $r \ge 0.06$. 
The existence of such a limit follows  
since the modifications to the standard consistency equation 
become smaller at lower energies, corresponding to $\sin \theta \rightarrow 
1$ in Eq. (\ref{tensortilt}). 

\begin{figure}[!t]
\includegraphics[width=7.2cm]{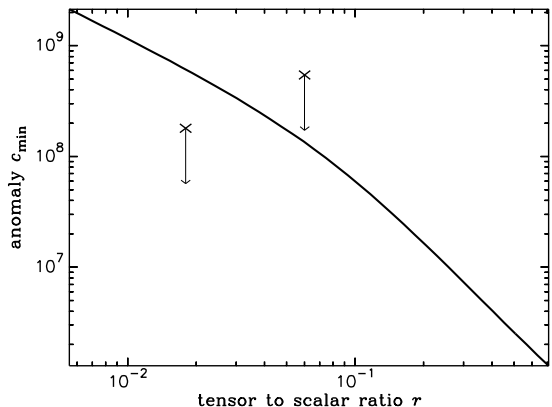}

\caption[] {\label{fig2holography}
As in Fig. \ref{fig1holography}, but illustrating the 
lower limit on the conformal anomaly coefficient 
that leads to a detectable departure from the standard 
inflationary consistency equation for a 
given observed tensor-to-scalar ratio.
}
\end{figure}

\section{Expectations from AdS/CFT}

Thus far, we have considered empirical constraints 
that need to be satisfied for a detection of Eq. (\ref{equivconsist}).
However, the conformal anomaly coefficient, $c$, 
is related to the departure parameter through Eq. (\ref{defrl})
and increases monotonically with $r_l$ for a given value of $r$. 
Hence, the necessary condition on $r_l$ for a detection of the 
holographic consistency equation may be interpreted as a necessary lower limit 
on the conformal anomaly coefficient, or equivalently, 
on the number of colours in the dual gauge theory. These limits  
are summarized in the fourth and fifth columns of Table I 
and Fig. \ref{fig2holography} 
\footnote{As a first approximation, we neglect 
any observational errors in $r_l$.}.  

Such limits may be compared to
what is expected from field theoretic considerations. 
In its simplest form, the AdS/CFT correspondence relates string theory 
on $AdS_5 \times S^5$ with $N$ units of four-form flux 
to ${\cal{N}} =4$ ${\rm SU} (N)$ superconformal field theory \cite{adscft}. 
For ${\rm AdS}_5$ backgrounds arising from compactifications of the 
type IIB string theory that include ${\rm D3}$-branes, 
$N$ represents the number of branes that are present. 
In the majority of string 
compactifications that have been studied to date, 
$N \approx 10$ and this is clearly well below detectable levels. 
Nonetheless, higher values of $N$ are possible \cite{higherN,gubser}.   
A powerful geometric way of describing compactifications 
of type IIB string theory is through F-theory compactification on 
Calabi--Yau four-folds, $K_8$, that admit an elliptic fibration with a section, 
i.e., are locally the product $K_8=K_6\times T^2$, where $K_6$ 
is a complex three-fold \cite{fourfold,eulerK8}. 
(The four-fold $K_8$ is not the physical 
compactified space, but provides a convenient way of parametrizing the 
geometry and moduli field VEVs). 

Global tadpole cancellation 
implies that the effective total ${\rm D3}$-brane charge is constrained by 
the topology of the $K_8$ such that $N_{D3} = \chi /24 -{\cal{F}}$, 
where $\chi (K_8)$ is the Euler characteristic of the four-fold 
and ${\cal{F}}$ represents the number of fluxes present 
\cite{fourfold,eulerK8}. 
Large values of $N$ are therefore possible for suitable 
choices of $K_8$. The topological properties of Calabi-Yau four-folds 
have been studied extensively \cite{fourfold,eulerK8}. 
For the case of manifolds that are known explicitly (corresponding to 
the class of four-folds that can be represented as 
hypersurfaces in weighted projective spaces) the 
Euler number can be as high as $\chi \le 1,820,448$ \cite{eulerK8}, 
thus allowing as many as $N_{D3} \approx \chi (K_8)/24 \approx 7.5 \times 10^4$ 
${\rm D3}$-branes to be present. This implies a maximal central charge 
of $c\approx 9 \times 10^6$ and comparing this 
with the minimal values presented in Table I indicates
that the scalar-to-tensor ratio must exceed 
$r>0.3$ if the holographic consistency equation is to be detectable. 
On the other hand, such a model would lack fluxes and 
would therefore probably not be realistic from a phenomenological 
point of view since the moduli fields would be effectively massless
\footnote{We thank E. Kiritsis for highlighting this point.}. 
Consequently, a conservative (and more realistic) 
estimate for the maximal value of the ${\rm D3}$-brane charge 
might be in the range $N_{D3} \approx 1-5 \times 10^4$, 
although it should be noted that a definitive mechanism 
for moduli stabilization has yet to be proposed.  
It should also be emphasized that the above discussion applies only for 
known manifolds and examples admitting a
larger ${\rm D3}$-brane charge are not ruled out at the present time. 
In this sense, therefore, $N$ may still be regarded as a free parameter. 

Such a large hidden CFT sector 
with $10N^2 \approx 10^{11}$ degrees of freedom could be problematic 
for cosmology, and primordial nucleosynthesis in particular. 
However, the nucleosynthesis bounds can be satisfied if 
the standard model and CFT degrees of freedom are decoupled and satisfy 
the condition $\rho_{\rm CFT} < \rho_{\rm CMB}$ at the present epoch, 
where $\rho_{\rm CMB}$ is the energy density of the CMB. 
This ensures that $\rho_{\rm CFT} < \rho_{\rm SM}$  at energy scales 
$\approx 1\, {\rm MeV}$, where a subscript `SM' denotes standard model 
degrees of freedom. This bound may not necessarily be 
satisfied if energy leaks from the standard model into the CFT, but  
it can be shown on dimensional grounds \cite{gubser} that this 
is not a problem if the parameter $\kappa \equiv \ell^2T^5_{\rm SM}H^{-1}l_P^2$
is less than unity, where $T_{\rm SM}$ denotes 
the temperature. Recalling that $H\propto a^{-3/2}$ and 
$H \propto a^{-2}$ during the matter- and radiation-dominated 
phases of the universe's history, respectively, and defining the 
redshift $z \equiv 1+a_0/a$, then implies that 
\begin{equation}
\ell \approx  \frac{\kappa^{1/2}}{l_P T_0^{5/2}
H_0^{-1/2} z_{\rm LSS}^{1/4}}\frac{1}{z^{3/2}}
\approx 6 \times 10^{48} \left( \frac{\kappa}{z^3} 
\right)^{1/2} l_P  \,
\end{equation}
where in the second equality we have taken $T_0 = 2.4 \times 10^{-13} 
{\rm GeV}$. Gravity waves generated from standard, 
single-field inflation will be detectable from polarization of the 
CMB if the energy scale of inflation exceeds $3.2 \times 10^{15}\, {\rm GeV}$, 
corresponding to an inflationary redshift of $z_{\rm inf} \approx
10^{28}$ \cite{othergravy}. Thus, requiring  
$\kappa < 1$ back to a redshift of $z_{\rm inf}$ then leads to 
an upper limit on the ${\rm AdS}_5$ length scale such that 
$\ell < 6 \times 10^6 l_P$. In the AdS/CFT dictionary, 
$N \approx \ell /l_P$, so this is comfortably within the bound required 
for detectability of the holographic consistency equation. 

\section{Conclusion}

At the present time, upper bounds on 
the scalar-to-tensor ratio are deduced from observations
of the CMB anisotropy power spectrum and high redshift surveys. 
These bounds are sensitive to the priors assumed on the 
value and running of the scalar spectral index, $n_S$, as well as the data 
sets included. The limit from the WMAP data alone, assuming no 
priors on $d n_S/d \ln k$ and $n_S$, is $r< 0.7$ at $95 \%$ confidence, but 
this strengthens to $r<0.49$ if no running in $n_S$ is assumed \cite{WMAP}.
Combining CMB data sets with other large-scale structure surveys 
leads to the somewhat tighter bounds of $r<0.36$ 
with no prior on the running and $r< 0.27$ 
with the prior $dn_S/d \ln k =0$ assumed \cite{seljak}. 

We have found that a necessary condition 
for an observable detection of the holographic consistency equation 
(\ref{equivconsist}) is that $r \ge 0.06$. This is well within the 
level of sensitivity attainable with near-future CMB polarization experiments 
such as Planck \cite{planck} and Clover \cite{clover}. Within 
the context of the AdS/CFT correspondence, this lower bound 
transforms into a limit on the number of colours in the dual 
gauge theory, $N > 3 \times 10^5$. This is only a factor of 4 or so higher 
than the upper limit attainable in known compactifications 
of the type IIB string. Indeed, 
for values in the range $N \approx 6 \times 10^4$ or lower, 
a detection should be possible in principle if $r  \ge 0.3$ and this 
is tantalizingly close to the current upper limits on 
the tensor-to-scalar ratio. This implies that 
a positive detection of the holographic consistency equation 
would place a strong limit on the central charge of the dual CFT 
in the range $N \approx 10^4-10^5$ and consequently 
on the topology of the higher dimensions.

On the other hand, we may take a more 
phenomenological view and interpret the Friedmann 
equation (\ref{friedmann}) as an independent four-dimensional model. 
In this case, a detection of the holographic consistency equation 
could be interpreted as observational evidence for modifications 
to Einstein gravity at very high energy scales. Moreover, 
if modifications to Einstein gravity of this nature are ever to be 
observable, we should expect to detect a gravitational wave 
background in the relatively near future.
Conversely, observations that indicated   
$n_T+r/4.8 \ne 0$ and $r<0.06$ would rule out 
the scenario (\ref{friedmann}).  
This offers the promise of constraining 
gravitational physics at energies inaccessible to any other 
form of experiment and the observations are eagerly awaited. 

\acknowledgments
DS is supported by PPARC. We thank E. Kiritsis and A. Liddle  
for helpful communications and L. Knox and Y-S. Song 
for making the numerical results of their paper \cite{songknox}
available to us.

\end{document}